\documentclass[prb,twocolumn,aps,showpacs]{revtex4}
\usepackage{graphicx}%
\usepackage{dcolumn}
\usepackage{amsmath}
\usepackage{amssymb}
\usepackage{epsfig}

\begin{document}

\title{ Interaction strengths in cuprates from the
inelastic neutron-scattering measurements}


\author{Z.G. Koinov}
\address{Department of Physics and Astronomy,
University of Texas at San Antonio, San Antonio, TX 78249, USA}
\email{Zlatko.Koinov@utsa.edu} 

\begin{abstract}

 The $t-U-V-J$ model is used to describe the positions of the experimental
  peaks associated with commensurate and incommensurate
 structure of the magnetic susceptibility probed by
neutron scattering in cuprate compounds. Assuming that the
tight-binding form of the mean-field electron energy and the maximum
gap can be obtained by fitting the angle-resolved photoemission
spectroscopy (ARPES) data, we have determined the strengths of the
on-site repulsive interaction $U$, the spin-independent attractive
interaction $V$ and the spin-dependent antiferromagnetic interaction
$J$ from the positions of the commensurate and incommensurate peaks.
 \end{abstract}

\maketitle
% main text

 \textbf{Introduction.} It is evident from many
experiments\cite{NSS,Bi,Ar} that the spin excitation spectrum in the
superconducting state of cuprate compounds for low temperature and
low energy consists of incommensurate (IC) magnetic peaks at the
quartet of wavevectors
$\textbf{Q}_{\delta}=(\pi(1\pm\delta),\pi(1\pm\delta))$ where
$\delta$ represents the degree of incommensurability (the lattice
parameter $a=1$ and $\hbar=1$
 are set to unity, $\textbf{Q}_{\delta}$ corresponds to
$(0.5(1\pm\delta),0.5(1\pm\delta))$ in reciprocal lattice units,
r.l.u., and the total number of sites is $N$). $\delta$ decreases
with increasing the energy transfer $\omega$ and vanishes at energy
$\omega_{res}$. This commensurate peak is called a magnetic
resonance peak and it is centered at the
 antiferromagnetic wave vector $\textbf{Q}_{AF}=(\pi,\pi)$. It was
suggested\cite{Plas,Plas1,HC}
 that the existence of the resonance peak in Bi2212 samples is related to two strong interactions:
 an on-site repulsion $U$ which drives the system close to an antiferromagnetic
instability, and  a short-range antiferromagnetic interaction $J$
related to the fact that cuprates are oxides of copper doped with
various other atoms (doped antiferromagnetic Mott insulators). Since
d-wave superconductivity on the cuprate square lattice does not rule
out the existence of a spin-independent attractive interaction, we
naturally arrive at the idea that the $t-U-V-J$ model could be a
possible theoretical scenario which fits together three major parts
of the high-$T_c$ superconductivity puzzle of the cuprate compounds:
(i) it describes the opening of a d-wave pairing gap, (ii) it is
consistent with the fact that the basic pairing mechanism arises
from the antiferromagnetic exchange correlations, and (iii) it takes
into account the charge fluctuations associated with double
occupancy of a site which play an essential role in doped systems.

In the one-layer approximation, the Hamiltonian of the
two-dimensional $t-J-U-V$ model is:
\begin{eqnarray}&\widehat{H}=-\sum_{i,j,\sigma}t_{ij}\psi^\dag_{i,\sigma}\psi_{j,\sigma}
-\mu\sum_{i,\sigma}\widehat{n}_{i,\sigma}+U\sum_i
\widehat{n}_{i,\uparrow} \widehat{n}_{i,\downarrow}\nonumber\\&
-V\sum_{<i,j>\sigma\sigma'}\widehat{n}_{i,\sigma}\widehat{n}_{j,\sigma'}+J
\sum_{<i,j>}\overrightarrow{\textbf{S}}_i\textbf{.}\overrightarrow{\textbf{S}}_j,
\label{Hubb1}\end{eqnarray} where $\mu$ is the chemical potential.
The Fermi operator $\psi^\dag_{i,\sigma}$ ($\psi_{i,\sigma}$)
creates (destroys) a fermion on the lattice site $i$ with spin
projection $\sigma=\uparrow,\downarrow$ along a specified direction,
and $\widehat{n}_{i,\sigma}=\psi^\dag_{i,\sigma}\psi_{i,\sigma}$ is
the density operator on site $i$ with a position vector
$\textbf{r}_i$. The symbol $\sum_{<ij>}$ means sum over
nearest-neighbor sites.  The spin operator is defined by
$\overrightarrow{\textbf{S}}_i=\psi^\dag_{i,\sigma}
\overrightarrow{\sigma}_{\sigma\sigma'}\psi_{i,\sigma'}/2$, where
 $\overrightarrow{\sigma}$ is the vector formed by the
Pauli spin matrices $(\sigma_x, \sigma_y,\sigma_z)$. The terms in
(\ref{Hubb1}) represent the hopping of electrons between sites of
the lattice, their on-site repulsive interaction ($U>0$), the
attractive interaction between electrons on different sites of the
lattice ($V>0$) and the spin-dependent Heisenberg near-neighbor
interactions, respectively.

The gap equation in the case of d-wave pairing
\begin{equation}1=\frac{V_\psi}{2}\int^\pi_{-\pi}\int^\pi_{-\pi}
\frac{d\textbf{k}}{(2\pi)^2} \frac{d^2_\textbf{k}}
{\sqrt{\overline{\varepsilon}^2_\textbf{k}+\Delta^2_\textbf{k}}},\label{BCS!}\end{equation}
where $V_\psi=2V+3J/2$,
$E(\textbf{k})=\sqrt{\overline{\varepsilon}^2_\textbf{k}+\Delta^2_\textbf{k}}$,
and $\Delta_\textbf{k}=\Delta(\cos k_x-\cos k_y)/2$ provides a
relationship between the strengths of the $V$ and $J$ interactions.
The mean-field electron energy $\overline{\varepsilon}_\textbf{k}$
has a tight-binding form
$\overline{\varepsilon}_\textbf{k}=t_1\left(\cos k_x+\cos
k_y\right)/2+t_2 \cos k_x\cos k_y +t_3(\cos 2k_x+\cos
2k_y)/2+t_4(\cos 2k_x\cos k_y+\cos 2k_y\cos k_x)/2+t_5\cos 2k_x\cos
2k_y -\mu$ obtained by fitting the ARPES data with a chemical
potential $\mu$ and hopping amplitudes $t_i$ for first to fifth
nearest neighbors on a square lattice.  Since the spin
susceptibility and the two-particle Green's function share common
poles, we can adjust the parameters $U,V$ and $J$ in such a way that
the resonance peak $\omega_{res}$ is a solution of the
Bethe-Salpeter (BS) equations for the spin collective mode of the
Hamiltonian (\ref{Hubb1}) at a wave vector $\textbf{Q}_{AF}$. The
observed IC peaks are also poles of the spin susceptibility (or
solutions of the BS equations), and therefore, we can use any of the
IC resonances to obtain the exact strengths of the interactions. The
calculated strengths could be tested using the positions of the
other IC peaks observed on the same sample but at different transfer
energy.

\textbf{Bethe-Salpeter equations for the collective modes.} Thought
the method  for solving the BS equation is not new,\cite{KN}  we
tread the subject in detail for the sake of completeness. The
antiferromagnetic spin-dependent interaction $J
\sum_{<i,j>}\overrightarrow{\textbf{S}}_i\textbf{.}\overrightarrow{\textbf{S}}_j=J_1+J_2$
consists of two terms:
$J_1=\frac{J}{4}\sum_{<i,j>}[\widehat{n}_{i,\uparrow}\widehat{n}_{j,\uparrow}
+\widehat{n}_{i,\downarrow}\widehat{n}_{j,\downarrow}-
\widehat{n}_{i,\uparrow}\widehat{n}_{j,\downarrow}-
\widehat{n}_{i,\downarrow}\widehat{n}_{j,\uparrow}]$ and
$J_2=\frac{J}{2}\sum_{<i,j>}\left[\psi^\dag_{i,\uparrow}\psi_{i,\downarrow}
\psi^\dag_{j,\downarrow}\psi_{j,\uparrow}
+\psi^\dag_{i,\downarrow}\psi_{i,\uparrow}
\psi^\dag_{j,\uparrow}\psi_{j,\downarrow}\right]$. The interaction
described by the $J_2$ term does not allow us to use two-component
Nambu fermion fields, and therefore,  we introduce four-component
Nambu fermion fields $\widehat{\overline{\psi}}
(y)=\left(\psi^\dag_\uparrow(y)\psi^\dag_\downarrow(y)\psi_\uparrow(y)\psi_\downarrow(y)
\right)$  and
$\widehat{\psi}(x)=\left(\psi^\dag_\uparrow(x)\psi^\dag_\downarrow(x)\psi_\uparrow(x)
\psi_\downarrow(x)\right)^T$, where $x$ and $y$ are composite
variables and the field operators obey anticommutation relations.
The "hat" symbol over any quantity $\widehat{O}$ means that this
quantity is a matrix.

  The interaction
part of the extended Hubbard Hamiltonian is quartic in the Grassmann
fermion fields so the functional integrals cannot be evaluated
exactly.  However, we can transform the quartic terms to a quadratic
form by applying the Hubbard-Stratonovich transformation for the
electron operators:\cite{ZK1} \begin{equation}\begin{split}&\int DA
e^{\left[\frac{1}{2}A_{\alpha}(z)D_{\alpha\beta}^{(0)-1}(z,z')A_{\beta}(z)+\widehat{\overline{\psi}}
(y)\widehat{\Gamma}^{(0)}_{\alpha}(y;x|z)\widehat{\psi}(x)A_{\alpha}(z)
\right]} \\&=e^{-\frac{1}{2}\widehat{\overline{\psi}}
(y)\widehat{\Gamma}^{(0)}_{\alpha}(y;x|z)\widehat{\psi}(x)
D_{\alpha\beta}^{(0)}(z,z') \widehat{\overline{\psi}}
(y')\widehat{\Gamma}^{(0)}_{\beta}(y';x'|z')\widehat{\psi}(x')}.\label{HSTr}\end{split}\end{equation}
The last equation is used to define the $4\times 4$ matrices
$\widehat{D}_{\alpha\beta}^{(0)}$ and
$\widehat{\Gamma}^{(0)}_{\alpha}$ ($\alpha,\beta=1,2,3,4$). Their
Fourier transforms, written in terms of the Pauli  $\sigma_i$, Dirac
$\gamma^0$ and alpha\cite{Y,B} matrices, are as follows:
$\widehat{D}^{(0)}=\left(\begin{array}{cc}\widehat{D}_1&0\\0&\widehat{D}_2\end{array}%
\right)$, $\widehat{\Gamma}_{1,2}^{(0)}=(\gamma^0\pm\alpha_z)/2$ and
 $\widehat{\Gamma}_{3,4}^{(0)}=(\alpha_x\pm \imath\alpha_y)/2$, where
$\alpha_i=\left(\begin{array}{cc}\sigma_i&0\\0&\sigma_y\sigma_i\sigma_y
\end{array}%
\right)$,  $\widehat{D}_1=
\left(J(\textbf{k})-V(\textbf{k})\right)\sigma_0+\left(U-J(\textbf{k})-V(\textbf{k})\right)
\sigma_x$ and $\widehat{D}_2=2J(\textbf{k})\sigma_x$.   For a square
lattice and nearest-neighbor interactions
$V(\textbf{k})=4V(\cos(k_x)+\cos(k_y))$ and
$J(\textbf{k})=J(\cos(k_x)+\cos(k_y))$.   After applying
transformation (\ref{HSTr}) the system under consideration consists
of a four-component boson field $A_{\alpha}(z)$ interacting with
fermion fields $\widehat{\overline{\psi}} (y)$ and
$\widehat{\psi}(x)$. The action of the model system is $S=
S^{(e)}_0+S^{(A)}_0+S^{(e-A)}$ where:
$S^{(e)}_0=\widehat{\overline{\psi}
}(y)\widehat{G}^{(0)-1}(y;x)\widehat{\psi} (x)$, $
S^{(A)}_0=\frac{1}{2}A_{\alpha}(z)D^{(0)-1}_{\alpha
\beta}(z,z')A_{\beta}(z')$ and $ S^{(e-A)}=\widehat{\overline{\psi}}
(y)\widehat{\Gamma}^{(0)}_{\alpha}(y,x\mid z)\widehat{\psi}
(x)A_{\alpha}(z)$. Here, we have introduced composite variables
$x,y,z=\{\textbf{r}_i,u\}$, where $\textbf{r}_{i}$ is a lattice site
vector, and  variable $u$ range from $0$ to $\beta=1/k_BT$ ($T$ and
$k_B$  are the temperature and the Boltzmann constant). We  use the
summation-integration convention: that repeated variables are summed
up or integrated over.

The idea of using the  Hubbard-Stratonovich transformation is to
establish an one-to-one correspondence between the $t-U-V-J$ model
and the model used to describe the Bose-Einstein condensate of
excitons in semiconductors\cite{GFS}. This allows us to follow the
same steps as in Refs. [\onlinecite{GFS}], and to derive a set of 20
BS equations for the collective mode $\omega(\textbf{Q})$ at zero
temperature. The Fourier transforms of $V$ and $J$ interactions are
separable, and therefore, we obtain a set of 20 coupled linear
homogeneous equations for the dispersion of the collective
excitations. The existence of a non-trivial solution requires that
the secular determinant $det\|\widehat{\chi}^{-1}-\widehat{V}\|$ is
equal to zero, where the bare mean-field-quasiparticle response
function $\widehat{\chi}=\left(
\begin{array}{cc}
P&Q\\
Q^T&R
\end{array}%
\right)$  and the interaction
$\widehat{V}=diag(U,U,-(U-2J(\textbf{Q})),U-2V(\textbf{Q}),-(2V+J/2),...,-(2V+J/2),-(2V-J/2),...,-(2V-J/2))$
are $20\times 20$ matrices. $P$ and $Q$ are $4\times 4$ and $4\times
16$ blocks, respectively, while $R$ is $16\times 16$ block (in what
follows $i,j=1,2,3,4$):
\begin{widetext}
  \begin{equation} P=\left|
\begin{array}{cccc}
I_{\gamma,\gamma}&J_{\gamma,l}&I_{\gamma,\widetilde{\gamma}}&J_{\gamma,m}\\
J_{\gamma,l}&I_{l,l}&J_{l,\widetilde{\gamma}}&I_{l,m}\\
I_{\gamma,\widetilde{\gamma}}&J_{l,\widetilde{\gamma}}&
I_{\widetilde{\gamma},\widetilde{\gamma}}&
J_{\widetilde{\gamma},m}\\
J_{\gamma,m}&I_{l,m}&J_{\widetilde{\gamma},m}&I_{m,m}
\end{array}%
\right|,Q=\left|
\begin{array}{cccc}
I^i_{\gamma,\gamma}&J^i_{\gamma,l}&I^i_{\gamma,\widetilde{\gamma}}&J^i_{\gamma,m}\\
J^i_{\gamma,l} &I^i_{l,l}&J^i_{l,\widetilde{\gamma}}&I^i_{l,m}\\
I^i_{\gamma,\widetilde{\gamma}}& J^i_{l,\widetilde{\gamma}}&I^i_{\widetilde{\gamma},\widetilde{\gamma}}&J^i_{\widetilde{\gamma},m}\\
J^i_{\gamma,m}&I^i_{l,m} &J^i_{\widetilde{\gamma},m} &I^i_{m,m}
\end{array}%
\right|, R=\left|
\begin{array}{cccc}
I^{ij}_{\gamma, \gamma}&J^{ij}_{\gamma, l}& I^{ij}_{\gamma,
\widetilde{\gamma}}&J^{ij}_{\gamma,
 m}\\
J^{ij}_{\gamma, l} &I^{ij}_{l,l}&
J^{ij}_{l,\widetilde{\gamma}}&I^{ij}_{l,m}\\
 I^{ij}_{\gamma
\widetilde{\gamma}}& J^{ij}_{l,\widetilde{\gamma}}&
 I^{ij}_{\widetilde{\gamma},\widetilde{\gamma}}&
J^{ij}_{\widetilde{\gamma}, m}\\
 J^{ij}_{\gamma,
 m}&I^{ij}_{l,m} &J^{ij}_{\widetilde{\gamma}, m}
  &I^{ij}_{m,m}
\end{array}%
\right|.\nonumber\end{equation}  The quantities
$I_{a,b}=F_{a,b}(\varepsilon (\mathbf{k},\mathbf{Q}))$ and
$J_{a,b}=F_{a,b}(\omega)$, the $1\times 4$ matrices
$I^i_{a,b}=F^{i}_{a,b}(\varepsilon (\mathbf{k},\mathbf{Q}))$ and
$J^i_{a,b}=F^{i}_{a,b}(\omega)$, and the $4\times 4$ matrices
$I^{ij}_{a,b}=F^{ij}_{a,b}(\varepsilon (\mathbf{k},\mathbf{Q}))$ and
$J^{ij}_{a,b}=F^{ij}_{a,b}(\omega)$ are defined as follows (the quantities $a(\mathbf{k}%
,\mathbf{Q})$ and $b(\mathbf{k},\mathbf{Q})=l_{\mathbf{k},\mathbf{Q}},m_{%
\mathbf{k},\mathbf{Q}},\gamma _{\mathbf{k},\mathbf{Q}}$ or $\widetilde{%
\gamma }_{\mathbf{k},\mathbf{Q}}$):
$$
F_{a,b}(x)\equiv \frac{1}{N}\sum_\textbf{k}\frac{%
xa(\mathbf{k},\mathbf{Q})b(\mathbf{k},%
\mathbf{Q})}{\omega ^{2}-\varepsilon ^{2}(\mathbf{k},\mathbf{Q})},
F^i_{a,b}(x)\equiv \frac{1}{N}\sum_\textbf{k}\frac{%
xa(\mathbf{k},\mathbf{Q})b(\mathbf{k},%
\mathbf{Q})\widehat{\lambda}^i_\textbf{k}}{\omega ^{2}-\varepsilon
^{2}(\mathbf{k},\mathbf{Q})},F^{ij}_{a,b}(x)\equiv \frac{1}{N}\sum_\textbf{k}\frac{%
xa(\mathbf{k},\mathbf{Q})b(\mathbf{k},%
\mathbf{Q})}{\omega ^{2}-\varepsilon
^{2}(\mathbf{k},\mathbf{Q})}\left(\widehat{\lambda}^T_\textbf{k}
\widehat{\lambda}_\textbf{k}\right)_{ij}.$$ \end{widetext} Here
$\varepsilon(\textbf{k},\textbf{Q})=
E(\textbf{k}+\textbf{Q})+E(\textbf{k})$, and the form factors are:
$\gamma_{\textbf{k},\textbf{Q}}=u_{\textbf{k}}u_{\textbf{k}+\textbf{Q}}+v_{\textbf{k}}v_{\textbf{k}+\textbf{Q}},\quad
l_{\textbf{k},\textbf{Q}}=u_{\textbf{k}}u_{\textbf{k}+\textbf{Q}}-v_{\textbf{k}}v_{\textbf{k}+\textbf{Q}},
\quad
\widetilde{\gamma}_{\textbf{k},\textbf{Q}}=u_{\textbf{k}}v_{\textbf{k}+\textbf{Q}}-u_{\textbf{k}+\textbf{Q}}v_{\textbf{k}},
$ and $ m_{\textbf{k},\textbf{Q}}=
u_{\textbf{k}}v_{\textbf{k}+\textbf{Q}}+u_{\textbf{k}+\textbf{Q}}v_{\textbf{k}}
$ where $u^2_{\textbf{k}}=1-v^2_{\textbf{k}}=
\left[1+\overline{\varepsilon}(\textbf{k})/E(\textbf{k})\right]/2$.

We can compare our BS equations with the previous approaches. If one
takes into account only the spin channel, the secular determinant is
a $1\times 1$ determinant and the equation for the collective mode
assumes the form
$1+(U-2J(\textbf{Q})I_{\widetilde{\gamma}\widetilde{\gamma}}=0$.
This is the well-known equation for the poles of  the spin
susceptibility  in the random phase approximation (RPA)
$\chi(\textbf{Q};\omega)=\chi^{(0)}(\textbf{Q};
\omega)/\left[1+(U-2J(\textbf{Q})\chi^{(0)}(\textbf{Q};\omega)\right]$,
where the BCS susceptibility at zero temperature is
$\chi^{(0)}(\textbf{Q};
\omega)=I_{\widetilde{\gamma}\widetilde{\gamma}}$.\cite{BS}
Obviously, the RPA overestimates spin fluctuations because the
mixing between the spin channel and other channels is neglected. In
the $m-$channel response theory we take into account the mixing
between the spin channel and $m-1$ other channels. The secular
determinant is $m\times m$. The coupling of the spin and two $\pi$
channels leads to a three-channel response-function theory.\cite{HC}
 The four-channel theory\cite{Plas,Plas1} includes the extended spin
channel to the previous three channels.  To see the difference
between the previous theories and the present approach, we rewrite
the secular determinant as
$det|\widehat{\chi}^{-1}-\widehat{V}|=det\left|
\begin{array}{cc}
A&B\\
B^T&C
\end{array}%
\right|=det|C|det|A-BC^{-1}B^T|$. The $m-$channel response-function
theory takes into account only the  $m\times m$ matrix A, neglecting
the mixing with the other $20-m$ channels which is represented by
the $m \times m$ matrix $BC^{-1}B^T$.

\textbf{Strengths of interactions in $Bi2212$ samples.} For Bi2212
compound, there are two possible sets of parameters with all
tight-binding basis functions involved (see Table 1 in Ref.
[\onlinecite{N}]). Assuming $\Delta=35$ meV, we obtain
$V^{(1)}_\psi=115.2$ meV with set 1, and
 $V^{(2)}_\psi=87.9$ meV with set 2.
Hao and Chubukov\cite{HC} have
 used another set of parameters (we shall call it H$\&$C) for Bi2212
compound with a doping concentration $x=0.12$: $t_1=-4t, t_2 =1.2t$,
$t=0.433$ eV, $\mu=-0.94t$, $\Delta=35$ meV and $V_\psi=0.6t$. The
parameters $U,V$ and $J$ should be adjusted in such a way that the
sharp collective mode of 40 meV which appears at wave vector
$\textbf{Q}_{AF}$ corresponds to the lowest collective mode
calculated by the BS equations.
\begin{figure}[tbp]
\includegraphics{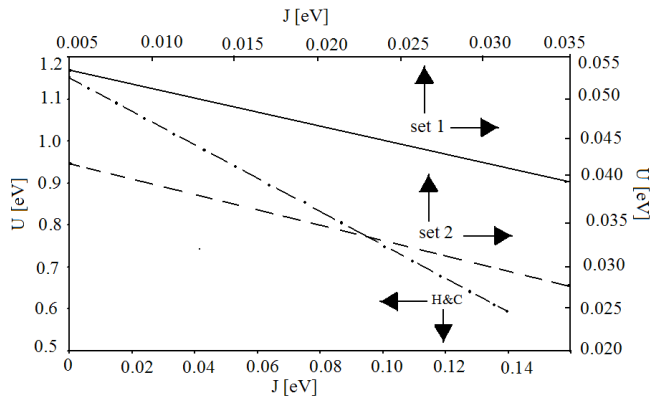} \label{Fig. 1}
\caption{Lines in $U,J$ parameter space which reproduce the
resonance energy of 0.04 eV. The H$\&$C line has been calculated
using the set of parameters given in Ref. [\onlinecite{HC}], while
set 1 and set 2 are obtained by using the two sets of parameters
given in Ref. [\onlinecite{N}].}
\end{figure}
In Fig. 1 we present the results of our calculations of the lines in
$U,J$ parameter space which reproduce the resonance energy of 40 meV
using all twenty channels but with set 1, set 2 and H$\&$C
parameters. Perhaps due to the limited size of single crystals
currently available, no incommensurate peaks in Bi2212 samples have
been reported so far. To obtain the exact values of the strengths we
need at least one IC peak.

\textbf{Strengths of interactions in $YBa_2Cu_3O_{6.7}$  samples.}
 To obtain the strengths of the interactions we use the
 commensurate peak at $\sim 40$ meV and incommensurate
peaks at $\sim 24$ meV and $\sim 32$ meV  which have been reported
in underdoped $YBa_2Cu_3O_{6.7}$ \cite{Ar}. It is known that the
YBaCuO is a two-layer material, but most of the peak structures
associated with the neutron cross section can be captured by one
layer band calculations \cite{DD}. The effects due to the two-layer
structure can, in principle, be incorporated in our approach, but
this will make the corresponding numerical calculations much more
complicated.

 The commensurate and incommensurate peak structures
associated with the neutron cross section in YBaCuO have been
studied within the electron-hole scenario using the single-band
Hubbard $t-U$ model \cite{Mc,Sch,Er} or the $t-J$ model
\cite{MW,BL,BL1,Li,Li1,YM}. The techniques that have been used are
based on (i) the Monte Carlo numerical calculations \cite{Mc}, (ii)
the random phase approximation (RPA) for the magnetic susceptibility
\cite{Sch,Er}, (iii) the mean-field approximation \cite{MW,BL} and
(iv) the RPA combined with the slave-boson mean field scheme
\cite{BL1,Li,Li1,YM}. It is known that in the case when the Hubbard
repulsion is large ($U/t \rightarrow \infty$),  the
antiferromagnetic exchange $J$ interaction is the consequence of
Hubbard repulsion, because the $t-J$ model is obtained after
projecting out the doubly occupied states in the Hubbard $t-U$
model, so that $J=2t^2/U$. Strictly speaking, by projecting out the
doubly occupied states we remove the high-energy degrees of freedom
and replace them with kinematical constraints assuming that the high
energy scale (given by $U$ that in cuprates corresponds to the
energy cost to doubly occupy the same site) is irrelevant. Thus, if
the constraint of no double occupancy is released, we arrive to the
conclusion that the the magnetic susceptibility should be calculated
using the $t-U-J$ model rather than the $t-U$ and $t-J$ models (the
corresponding arguments are presented in Refs.
[\onlinecite{K,J,K1,X}]).

  In our calculations the
mean-field electron energy $\overline{\varepsilon}_\textbf{k}$ has a
tight-binding form $\overline{\varepsilon}_\textbf{k}=-2t\left(\cos
k_x+\cos k_y\right)+4t' \cos k_x\cos k_y -2t''(\cos 2k_x+\cos
2k_y)-\mu$ . Using the established approximate parabolic
relationship  $T_c /T_{c,max}=1-82.6(p-0.16)^2$, where $Tc,max\sim
93$K is the maximum transition temperature of the system, $T_c=67$K
is the transition temperature for underdoped $YBa_2Cu_3O_{6.7}$, we
find that  the hole doping is $p=0.10$. At that level of doping the
ARPES parameters are obtained in Refs. [\onlinecite{ARPES}]:
$t=0.25$ eV, $t'=0.4t$, $t''=0.0444t$ and $\mu=-0.27$ eV.  In the
case of d-pairing the gap function is $\Delta_\textbf{k}=\Delta
d_\textbf{k}/2$, where the  gap maximum $\Delta$ should agree with
ARPES experiments. In the case of underdoped $YBa_2Cu_3O_{6.7}$ the
gap maximum has to be between the corresponding $\Delta=66$ meV in
$YBa_2Cu_3O_{6.6}$ and $\Delta=50$ meV in $YBa_2Cu_3O_{6.95}$
\cite{Delta}, so we set $\Delta=60$ meV. The numerical solution of
the gap equation provides $V_\psi=265$ meV. Next, we have solved
numerically the BS equations to obtain the spectrum of the
collective modes $\omega(\textbf{Q})$ at the commensurate point
$\textbf{Q}_{AF}$, as well as at four incommensurate points
$\pi(1\pm \delta),\pi)$ and $\pi,\pi(1\pm \delta))$.  From the 40
meV solution of the BS equations at $\textbf{Q}_{AF}$ we obtain a
relation between $U$ and $J$ parameters, which is represented by the
linear formula $U=-3.985 J+1.01$ ($U$ and $J$ are in eV). By means
of the last relation we have solved the BS equations for the one of
the four incommensurate 24 meV peaks ($\delta=0.22$). \cite{Ar} The
solution provides the following interaction strengths: $J\sim 129$
meV, $V\sim 35.7$ meV and $U\sim 495$ meV. To test the above values
of the strengths we calculated the positions of the incommensurate
peaks at 32 meV ($\delta=0.19$).\cite{Ar} The BS equations with the
above strengths provide the deviation from $\textbf{Q}_{AF}$ of
about $\delta=0.192$, which is in excellent agreement with the
experimentally obtained deviation (see FIG. 2 in Ref.
[\onlinecite{Ar}]). The strength of $J$ is in a very good agreement
with the strength of the superexchange interactions in the
underdoped antiferromagnetic insulator state of the cuprates which
in YBaCO family has a magnitude of $0.1-0.12$ eV, though some
theoretical papers have predicted similar magnitudes.

\textbf{Strengths of interactions in $La_{2-0.16}Sr_{0.16}CuO_4 $
samples.} To obtain the strengths of the interactions we use two IC
 peaks at $\sim 7.1$ meV ($\delta=0.261$) and at $\sim 10$ meV ($\delta=0.255$) which have been reported
in Refs. [\onlinecite{CH}]. The tight-binding band parameters
interpolated from values given in Refs. [\onlinecite{Tbp}]: $t=0.25$
eV, $t'=0.148t$, $t''=-0.5t$ and $\mu=-0.821t$. The maximum energy
gap (estimated according to the prediction of the mean-field
theory\cite{MFG}) is about 10 meV, and therefore, $V_{\psi}=95.8$
meV, which corresponds to maximum value of $J$ about 64 meV. The
solutions of the BS equations provide the following interaction
strengths: $J\sim 44.1$ meV, $V\sim 14.8$ meV and $U\sim 442.5$ meV.
To test these parameters we calculated the position of the  IC peak
at 18 meV.  The calculated value of $\delta=0.19$ is in very good
agreement with the experimental value of $\delta=0.186$.\cite{CH}
The above strengths provide the position of the commensurate peak to
be at $\sim 40$ meV, which is in a very good agreement with the
experimental value of $41\pm 2.5$ meV.\cite{CH} It is worth
mentioning that in $La_2CuO_4$ $J\sim 120$ meV, and therefore, our
calculations support the idea that the antiferromagnetic interaction
decreases with doping.\cite{DJ}\\
\textbf{Conclusion} Our  approach to the incommensurate and
commensurate structure of the magnetic susceptibility is based on
the conventional idea of particle-hole excitations around the Fermi
surface. There exists a second interpretation of the structure of
the magnetic susceptibility in terms of the spin-charge stripe
scenario (see Ref. [\onlinecite{Str}] and the references therein)
according to which the incommensurate peaks are natural descendants
of the stripes, which are complex patterns formed by electrons
confined to separate linear regions in the crystal. We do not wish
to repeat the theoretical arguments that were advanced against the
stripes model, but our unified description of the peaks based on the
$t-U-V-J$ model strongly supports the idea put forward by various
groups that the commensurate resonance and the incommensurate peaks
in cuprate compounds have a common origin.

\end{document}